# Synchronous Clock and RF Carrier Transmission for Radio Access Network Fronthaul

Kari A. Clark[(1)], Zun Htay[(1)], Zichuan Zhou[(1)], Amany Kassem[(1)], Andrea Pertoldi[(2)], Benjamin Rudin[(2)], Florian Emaury[(2)], Izzat Darwazeh[(1)], Zhixin Liu[(1)]

[(1)] Dept. of Electronic & Electrical Engineering, UCL, kari.clark.14@ucl.ac.uk
[(2)] Menhir Photonics AG, florian.emaury@menhir-photonics.com

**Abstract** *We simultaneously achieve clock synchronisation, clock-synchronised data transmission and ultra-low noise RF carrier generation by combining clock phase caching and frequency comb transmission in radio access networks (RAN). We demonstrate <100fs jitter for 25GHz RF carrier and 2.5GHz clock, and 16-hour 6.6ps RMS wander.*
©2025 The Author(s)

**Introduction**

6G mobile communications emphasizes emerging user cases including the integration of sensing and communications (ISAC), ultra-precise (cm) positioning, and high-speed radio wireless communications [1]. Precise positioning requires ps-level clock synchronisation between multiple radio units (RU). ISAC requires low noise and frequency synchronised RF carriers over distributed antennas for environmental sensing. High-capacity radio wireless communication requires low noise carriers over multiple bands into high frequency microwave/millimetre-wave bands [2-3].

Currently, the clock synchronisation and RF carrier generation/synchronisation are treated as separate problems, each addressed using separate devices and systems. For example, current clock synchronisation in radio access networks (RANs) uses Synchronous Ethernet (Sync-E) [4] and the precision time protocol (PTP) [5], which only achieves ns-level clock precision, insufficient for the demand for cm-level positioning that requires <30ps clock precision. Global navigation satellite systems (GNSS) can achieve ps-level timing accuracy but only with clear visibility of the open sky and is susceptible to spoofing and jamming from the very low signal power [6].

Traditional methods for generating RF signals, such as with electronic phase lock-loop (PLL), necessitate substantial silicon area and high-power consumption, rendering them non-scalable solutions for ultra-dense cellular networks. Further, as phase noise scales quadratically in conventional electronic PLLs, how to generate low phase noise and phase synchronised carriers over multiple RUs is a major challenge.

In this paper, we propose and demonstrate a unified approach to achieve ps-level clock synchronisation and synchronised low-noise RF carrier generation by combining clock phase caching, data and frequency comb transmission in a proof-of-concept RAN system.

Previously, we have demonstrated ps-level clock synchronisation over a point-to-point link using clock phase caching [7], only optical data and clock transmission was demonstrated. In separate work [3], we demonstrated low-noise phase synchronised RF carrier generation over multiples bands (sub-6GHz to 110GHz) by transmitting an optoelectronic frequency comb over fibre. However, the clock phase was not synchronised, so the system was unable to provide accurate positioning or wireless sensing. To the best of our knowledge, the combined ps-level clock synchronisation and ultra-low noise RF carrier transmission have not been demonstrated.

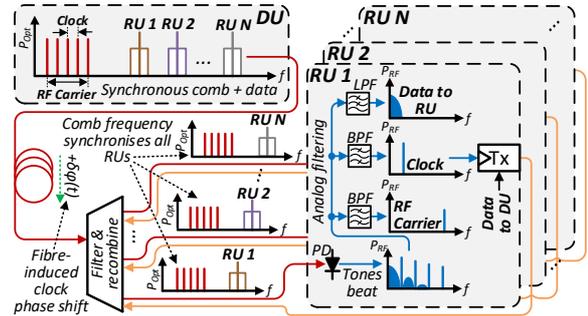

**Fig. 1:** Concept of optical frequency comb and data co-transmission from a distributed unit (DU) with co-reception at radio units (RU) by a single photodiode (PD) followed by analog filter separation into synchronous clock, RF carrier and data; $+\varphi(t)$, 1-way fibre clock phase shift, which we compensate with clock phase caching.

Here, we use a ultra-low noise frequency comb at the distributed unit as clock and frequency reference (see **Fig. 1**). By distributing an optical frequency comb and detecting with a photodetector (PD), we obtain 2.5 GHz as a clock reference and simultaneously multiple phase and frequency synchronised RF carrier tones at 2.5 GHz intervals up to the bandwidth limit of the PD. However, the optical fibre itself introduces up to ns of long-term clock phase drift from temperature change [9]. Traditionally, this is addressed by fibre noise cancellation using bi-directional transmission of clock or optical signals. Here we address this by compensation of long-term 1-way fibre clock drift in RANs by clock phase caching [5], providing low-cost and efficient clock stabilisation in a point-to-multipoint network.

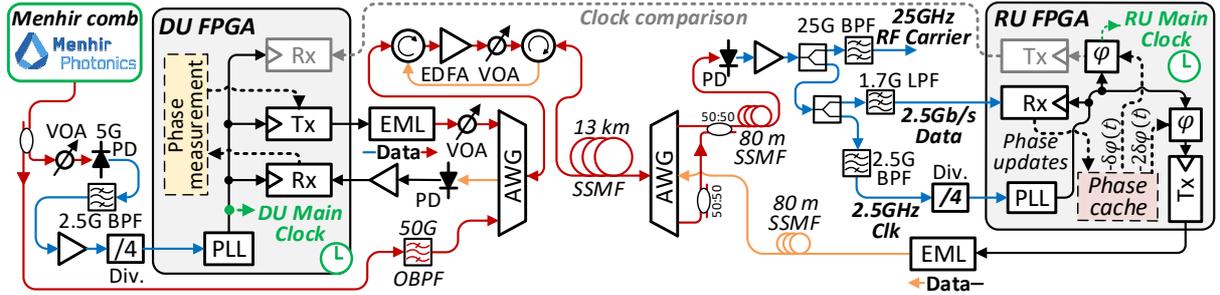

**Fig. 2:** Experimental setup to demonstrate our combined approaches to clock synchronisation in RANs.

Combining our approaches, we simultaneously demonstrate 6.6 ps root-mean-square (RMS) clock wander long-term stability and error free clock-synchronised transmission over 16 hours; and low phase noise 25-GHz RF carrier of <100 fs jitter (1KHz-10MHz), in a proof-of-concept RAN system.

**System architecture**
**Fig. 1** shows the conceptual diagram of the proposed clock and carrier synchronised RAN. An optical frequency comb (Menhir comb) with 2.5-GHz tone spacing is filtered and wavelength multiplexed with data signal at the distributed unit (DU) before being sent to the RUs. At the RU, the comb signal is demultiplexed, split and recombined with data signals before sending to each RU via feeder fibres. As a result, each photodiode (PD) will simultaneously detect both comb signal and data signal. With a comb-data wavelength separation higher than the PD bandwidth, the PD will generate an RF comb, with a spacing equal to that of the optical comb tones ($f_{rep}$) [8] and overlaid atop the data. Analog filtering is used to extract clock, data and RF carrier, enabling single PD reception for low-cost RUs.

As shown in **Fig. 1**, optical fibre introduces $+\delta\varphi(t)$ of slow clock phase drift from temperature change per direction (about 39 ps/(km·K) for standard single mode fibre (SSMF) [9]). With the RU upstream transmitter driven by the disseminated clock, we effectively build a clock return link that permits cancellation of fibre induced clock phase shift. The 2-way change in clock phase offset, $+2\delta\varphi(t)$, of data arriving from the RU at the DU is measured at the phase update rate of $f_\varphi$. Updates are sent to the RU at the same rate, which corrects the phase of its data sent to the DU by $-2\delta\varphi(t)$ at $f_\varphi$, cancelling the fibre clock phase shift of the RU to DU data. The RU also corrects for the 1-way clock shift, $+\delta\varphi(t)$, by halving the measured 2-way clock shift to $-\delta\varphi(t)$, cancelling the 1-way clock shift $+\delta\varphi(t)$ in the outbound path, in turn clock phase synchronising the RU's clock to the DU's clock.

**Experimental setup**
**Fig. 2** shows our experimental setup used to investigate the performance of our proposed approach. Limited by the available resources, we only developed one RU. At the DU side, we used a Menhir comb with 2.5 GHz $f_{rep}$ and centre frequency 1551.1 nm as the comb source. The Menhir comb features robustness in practical operation environment, compact and ultra-low phase noise that meets the system requirement. We split the comb at the DU, beat the tones on a 5 GHz PD, filter with a 2.5 GHz BPF, amplify, then used a /4 divider to generate a 625 MHz reference clock to synchronise the DU FPGA (field programmable gate array). The DU FPGA transmitter (Tx) generated real-time 2.5 Gb/s non-return-to-zero (NRZ) on-off-keyed (OOK) packet data containing $2^9$ length pseudorandom binary sequences (PRBS-9) sequences and medium access control (MAC) layer information that was modulated onto an 1554.13 nm carrier using an externally modulated laser (EML), attenuated to -19.75 dBm to minimise erbium doped fibre amplifier (EDFA) gain competition with the comb. The second split at the DU passed through an optical bandpass filter (OBPF) tuned to 50 GHz to minimise the negative impact of wavelength dispersion fading on the strength of the 25 GHz comb beat tone and its phase noise performance. The filtered comb, at -10.17 dBm, was combined with the synchronous data from the DU with a 200-GHz-spacing wavelength multiplexer before outbound-only amplification with an EDFA. The combined comb and data then launched at 10.35 dBm into 13 km SSMF. These signals were then demultiplexed by another 200-GHz wavelength demultiplexer. The comb was split by a 50:50 splitter to emulate a two RU system and then recombined with the data (-9.77 dBm) with a coupler. The recombined signals passed through 80 m of feeder fibre, before reception at -7.08 dBm on a 40 GHz PD. The signals were then amplified, split in the RF domain and the 25 GHz RF carrier, 2.5 GHz clock and 2.5 Gb/s data were filtered out with band pass filters (BPF) and a low pass filter (LPF) respectively. The 2.5 GHz clock was divided to 625 MHz with a /4 divider, which synchronised the reference clock of the RU FPGA. The RU FPGA generated return

-1.00 dBm data packets modulated onto a 1555.74 nm carrier which passed through the 13 km SSMF back to the DU before reception by an 18 GHz PD at -17.7 dBm. Clock phase caching was then implemented using the DU FPGA by measuring the $+2\delta\varphi(t)$ shift at its optical-link facing Rx every 0.1 s. Clock phase updates were sent through the optical link to the RU Rx. These updated a phase store which was used to drive a 1st phase interpolator (PI, $\varphi$) driving the return RU to DU data transmitter by $-2\delta\varphi(t)$ and a 2nd PI by only $-\delta\varphi(t)$, which drove the RU's main clock. To evaluate how well synchronised the RU was to the DU, this main RU clock drove a <30cm coax. link (grey) to the DU, the clock phase of which was measured by the DU at a rate of 9.54 kHz.

### Results and Discussion

We first investigate the phase noise of the 25 GHz RF carrier and 2.5 GHz clock generated at the RU from the optical frequency comb, as shown in **Fig. 3**. For the 25 GHz RF carrier, we show an overall similar jitter (integrated from 1 kHz to 10 MHz) of about 90 fs with and without co-reception of the data signals, indicating that the data clock tones have negligible impact of the RF carrier phase noise. For the 2.5 GHz clock, we show an overall integrated jitter over the same integration range of 70.3 fs without data. However, with data co-reception, the jitter degrades to 93.1 fs due to noise added from the data signals. The white noise floor in both cases is <-130 dBc/Hz. The eye diagram inset in **Fig. 5** shows that despite 6.57 dB greater optical clock power than data, the BER of a large proportion of the received eye still remains under $10^{-10}$, and we anticipate optimisation of analog filtering will result in further improvements. As a comparison against the standard commercial approach of synchronising RUs to the embedded clock in the data from the DU, we also measured the jitter of the 2.5 GHz embedded tone in PRBS-9 data packets generated from the DU FPGA to be 18 ps, over 2 orders of magnitude worse phase noise compared to our approach.

RoF comb clock distribution also provides the basis that allows clock phase caching to be possible. By frequency synchronising the RU using the comb, the clock phase drift is dominated by optical fibre induced clock phase drift. **Fig. 4** shows the 16-hour long-term stability resulting from running clock phase caching. The red (top) curve shows the DU-measured 2-way clock phase shift of data arriving at the DU from the RU through the optical fibre link, calculated from the phase updates sent to the RU. The blue (bottom) curve shows the RU clock phase synchronised to the DU. The green (middle) curve shows the estimated 1-way shift of the RU's clock that would have occurred without clock phase caching, calculated by adding half the measured 2-way clock phase offsets to the synchronised RU clock phase offset. Finally, **Fig. 5** shows the distribution of the recorded clock phase offset values between the RU and DU: showing 16-hour 6.61 ps RMS wander with clock phase caching, meeting the <30 ps requirement for cm-level positioning.

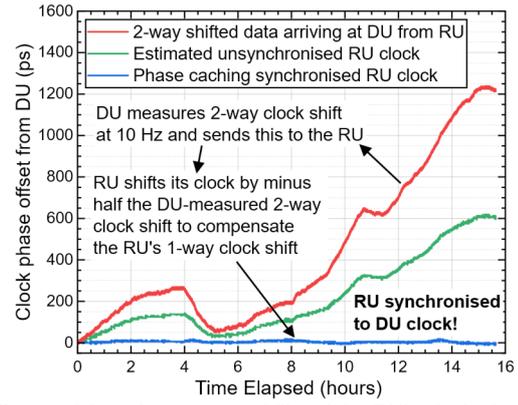

**Fig. 4:** 16-hour long-term stability of the RU's clock phase offset from the DU with clock phase caching.

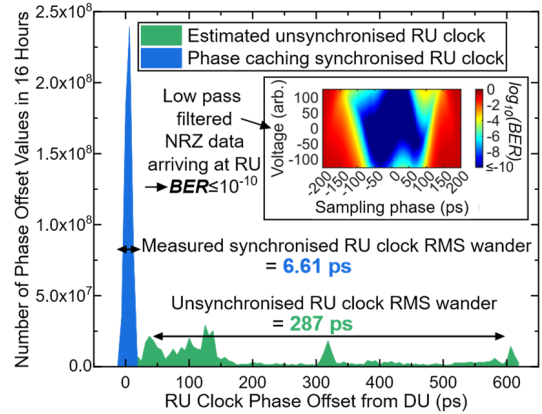

**Fig. 5:** Long-term RU clock phase stability using clock phase caching; *Inset*: error-free data after analog filtering.

### Conclusions

We demonstrate sub-100 fs jitter for both 2.5 GHz clock and 25 GHz RF carrier, error free data transmission and 6.61 ps RMS wander over 16 hours using clock phase caching. We allow a single high-quality optical oscillator to be shared by all RUs, side-stepping RU crystal oscillator trade-offs between cost, power and stability.

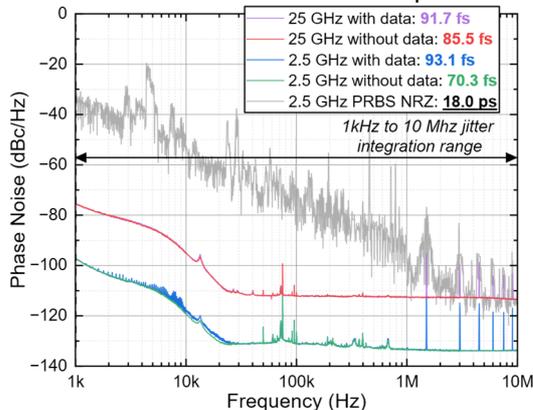

**Fig. 3:** Very low jitter of 25 GHz RF carrier and 2.5 GHz clock at RU after optical frequency comb and data reception.

**Acknowledgements**

Royal Academy of Engineering Research Fellowship (RF2122-21-234), Horizon Europe 6G MUSICAL (101139176), Innovate-UK USYNC (10089417), EPSRC TRANSNET (EP/R035342/1).